\documentclass[acmsmall]{acmart}

\usepackage{graphicx}

\usepackage{float}
\usepackage{caption}
\usepackage{subcaption}

\usepackage{rotating} 
\usepackage{textcomp}
\usepackage{xcolor}

\usepackage{multirow}
\usepackage{booktabs}

\usepackage{makecell}
\newcommand{\tabincell}[2]{\begin{tabular}{@{}#1@{}}#2\end{tabular}}

\usepackage{algorithm}
\usepackage{algpseudocode}

\usepackage{adjustbox}

\usepackage{ifthen}
\usepackage{xcolor} 
\newboolean{showcomments}
\setboolean{showcomments}{true}
\ifthenelse{\boolean{showcomments}}
  {
    \newcommand{\mynote}[2]{
        \fbox{\bfseries\sffamily\scriptsize#1}
        {%
          \small$\blacktriangleright$%
          \textsf{%
            \emph{%
              \ifthenelse{\equal{#1}{RS}}{\textcolor{red}{#2}}{%
                \ifthenelse{\equal{#1}{Brygon}}{\textcolor{blue}{#2}}{%
                  \ifthenelse{\equal{#1}{Zheng}}{\textcolor{orange}{#2}}{%
                    \textcolor{black}{#2}
                  }%
                }%
              }%
            }%
          }%
          $\blacktriangleleft$%
        }
    }
  }
  {
    \newcommand{\mynote}[2]{}
  }

\AtBeginDocument{%
  }

\setcopyright{acmlicensed}
\copyrightyear{2018}
\acmYear{2018}
\acmDOI{XXXXXXX.XXXXXXX}
\acmConference[Conference acronym 'XX]{Make sure to enter the correct
  conference title from your rights confirmation email}{June 03--05,
  2018}{Woodstock, NY}
\acmISBN{978-1-4503-XXXX-X/2018/06}

\begin{document}

\title{MBFL-DKMR: Improving Mutation-based Fault Localization through Denoising-based Kill Matrix Refinement}


\author{Hengyuan Liu}
\email{trovato@corporation.com}
\orcid{0000-0002-5884-2089}
\affiliation{%
  \institution{Beijing University of Chemical Technology}
  \city{Beijing}
  \country{China}
}

\author{Xia Song}
\affiliation{%
  \institution{Beijing University of Chemical Technology}
  \city{Beijing}
  \country{China}}
\email{2384513938@qq.com}

\author{Yong Liu}
\authornotemark[1]
\affiliation{%
  \institution{Beijing University of Chemical Technology}
  \city{Beijing}
  \country{China}}
\email{lyong@mail.buct.edu.cn}

\author{Zheng Li}
\authornote{Corresponding Authors}
\affiliation{%
  \institution{Beijing University of Chemical Technology}
  \city{Beijing}
  \country{China}}
\email{lizheng@mail.buct.edu.cn}

\begin{abstract}

Software debugging is a critical and time-consuming aspect of software development, with fault localization being a fundamental step that significantly impacts debugging efficiency. Mutation-Based Fault Localization (MBFL) has gained prominence due to its robust theoretical foundations and fine-grained analysis capabilities. However, recent studies have identified a critical challenge: noise phenomena, specifically the false kill relationships between mutants and tests, which significantly degrade localization effectiveness. While several approaches have been proposed to rectify the final localization results, they do not directly address the underlying noise.
In this paper, we propose a novel approach to refine the kill matrix, a core data structure capturing mutant-test relationships in MBFL, by treating it as a signal that contains both meaningful fault-related patterns and high-frequency noise. Inspired by signal processing theory, we introduce DKMR (Denoising-based Kill Matrix Refinement), which employs two key stages: (1) signal enhancement through hybrid matrix construction to improve the signal-to-noise ratio for better denoising, and (2) signal denoising via frequency domain filtering to suppress noise while preserving fault-related patterns. Building on this foundation, we develop MBFL-DKMR, a fault localization framework that utilizes the refined matrix with fuzzy values for suspiciousness calculation.
Our evaluation on Defects4J v2.0.0 demonstrates that MBFL-DKMR effectively mitigates the noise and outperforms the state-of-the-art MBFL techniques. 
Specifically, MBFL-DKMR achieves 129 faults localized at Top-1 compared to 85 for BLMu and 103 for Delta4Ms, with negligible additional computational overhead (0.11 seconds, 0.001\% of total time). This work highlights the potential of signal processing techniques to enhance the effectiveness of MBFL by refining the kill matrix.

\end{abstract}



\received{20 February 2007}
\received[revised]{12 March 2009}
\received[accepted]{5 June 2009}

\maketitle

\section{Introduction}
\label{sec:Introduction}  

Software testing and debugging are critical processes in the software development lifecycle, with significant implications for quality, reliability, and maintenance costs. In today's increasingly software dependent industries from healthcare systems managing patient data to financial platforms processing millions of transactions, the consequences of failures can be catastrophic, resulting in financial losses, compromised safety, or damaged reputation~\cite{britton2013reversible}. Software debugging consumes approximately 50-75\% of maintenance costs~\cite{wong2016survey}, with developers spending up to 35\% of their debugging time searching for fault locations~\cite{perscheid2017studying}, highlighting the importance of accurate and efficient fault localization techniques.

Among various fault localization techniques, Mutation-Based Fault Localization (MBFL) has emerged as a promising approach due to its theoretical foundations~\cite{papadakis2015metallaxis, moon2014ask} and fine-grained analysis capabilities~\cite{wang2025systematic}. MBFL leverages mutation testing to identify fault locations by analyzing the relationships between test outcomes and artificially introduced faults (i.e., mutants). Specifically, MBFL first generates mutants by creating syntactic variants of the original program, then executes the test suite on the mutants to collect and organize kill relationships between mutants and tests (often represented as a matrix), and finally computes suspiciousness scores based on these relationships. Unlike spectrum-based approaches that rely solely on coverage information, MBFL provides deeper insights into program behavior by utilizing rich diagnostic information from the kill matrix, which records the test results of each mutant, to examine how code modifications affect test outcomes~\cite{kim2023learning}. A variety of techniques have been explored to improve the effectiveness and efficiency of MBFL, including mutant generation~\cite{li2021hmbfl, du2024neural} and suspiciousness formula~\cite{wang2025systematic, du2022improving}. The kill matrix, serving as the foundation for suspiciousness calculation, has received particular attention. 
Researchers have explored different kill conditions to enrich the mutant-test information in the kill matrix~\cite{li2017transforming} and developed various learning-based techniques to reduce the kill matrix construction costs~\cite{kim2022predictive, kim2021ahead}.
Recent studies have observed \emph{noise} phenomena in MBFL~\cite{liu2024delta4ms, liu2025scope}, which mislead suspiciousness calculations and degrade localization accuracy.
Specifically, the program may contain false kill relationships between mutants and tests. 
The existence of these relationships can be attributed to testing inadequacy that fails to expose behavioral differences between mutants and the original program, incomplete error propagation where faults do not manifest as observable failures, etc.
Liu et al.~\cite{liu2024delta4ms} report that the noise from mutants can lead to biased suspiciousness scores. 
The causes of noise are diverse and complex. 
Directly analyzing whether an individual mutant constitutes noise requires examining extensive static or dynamic program behavior information, resulting in prohibitively high computational costs. In response to this challenge, a number of approaches have been proposed, including bias estimation for elimination~\cite{du2022improving,liu2024delta4ms} and statistical coupling analysis for mutant selection~\cite{jang2022hotfuz, kim2023learning}.
These approaches acknowledge the presence of noise, and they attempt to rectify the final results rather than remove the noise directly. This motivates us to explore more fundamental approaches that can address noise at its source.
%

The concept of \emph{noise} is inherently linked to the context of \emph{signals}. 
If the kill matrix can be regarded as a type of \emph{signals}, then the algorithms employed in signal processing can be utilized to address the \emph{noise} phenomenon in the MBFL.
A further investigation reveals that when the kill matrix is organized in an appropriate order, it can be viewed as a two-dimensional signal. 
When mutants are ordered by their code positions, the mutant dimension approximates a signal representing program behavior changes in test case data flow and control flow caused by modifications to different code entities. 
When test cases are ordered by test suites and functionality, the test case dimension approximates a signal representing the impact variations of mutant-injected faults under test inputs that test similar functionality but target different code entities. 
Moreover, MBFL's assumption that mutants from faulty statements should exhibit significantly higher kill rates compared to those from correct statements naturally leads to a frequency separation where meaningful fault signals manifest as low-frequency signal components while noise appears as high-frequency ones.
In light of the aforementioned findings, we hereby propose a novel Denoising-based Kill Matrix Refinement (DKMR) approach for MBFL. This approach refines kill matrices through signal processing techniques to mitigate the noise before suspiciousness calculation. 
The proposed approach is designed with two stages: 
(1) signal enhancement through hybrid matrix construction that combines weak and strong kill information to improve signal-to-noise ratio (SNR), making the subsequent denoising stage more effective, and 
(2) signal denoising through frequency domain filtering that suppresses high-frequency noise while preserving meaningful fault-related patterns. 
MBFL-DKMR is then developed as a fault localization framework that utilizes these refined matrices with fuzzy values for suspiciousness calculation, departing from traditional boolean-based approaches.

We evaluate MBFL-DKMR on Defects4J v2.0.0~\cite{gay2020defects4j}, a widely used benchmark containing real-world faults from diverse industrial projects. 
Our experiment results validate the effectiveness of DKMR in noise suppression through ablation studies, demonstrating that signal denoising successfully mitigates high-frequency noise while signal enhancement further improves SNR with statistical significance. 
Furthermore, when compared against state-of-the-art MBFL techniques including BLMu~\cite{wang2025systematic} and Delta4Ms~\cite{liu2024delta4ms}, MBFL-DKMR achieves better performance across all evaluation metrics, localizing 44 more faults at Top-1 than BLMu and 26 more than Delta4Ms. 
Additionally, the computational overhead analysis reveals that the kill matrix refinement process introduces negligible time cost, accounting for only 0.001\% of the total execution time.

We summarize the main contributions of our study as follows:

\begin{itemize}
    \item 
    \textbf{Signal View of Kill Matrix:} We establish a perspective that views kill matrix as signals containing meaningful fault-related patterns and high-frequency noise, which provides the foundation for applying signal processing techniques to MBFL. Furthermore, our analysis reveals the negative impact of high-frequency noise on fault localization effectiveness, necessitating effective denoising approaches.
    
    \item
    \textbf{DKMR: Denoising based Kill Matrix Refinement Method:} We propose DKMR, a novel kill matrix refinement approach that leverages signal processing techniques to filter the noise of the kill matrix. Two stages are designed: signal enhancement (combining strong and weak kill information to improve SNR to facilitate more effective denoising) and signal denoising (applying frequency domain filtering to suppress high-frequency noise).
    
    \item
    \textbf{MBFL-DKMR: Fault Localization Framework:} We develop MBFL-DKMR, an MBFL framework that utilizes a refined kill matrix from the DKMR method to perform suspiciousness calculation with fuzzy values, departing from traditional boolean-based approaches. This framework captures more nuanced relationships between mutants and test outcomes from the fuzzy value in the refined kill matrix and improves localization effectiveness.
    
    \item
    \textbf{Comprehensive Empirical Study:} Our extensive evaluation on Defects4J v2.0.0 validates the effectiveness of DKMR for improving fault localization accuracy, which includes ablation analyses of MBFL-DKMR, an assessment of its computational overhead, and a systematic comparison with state-of-the-art MBFL techniques across multiple metrics, including Top-N, MAP, and EXAM scores.
\end{itemize}


\section{Background}
\label{sec:Background}

\subsection{Mutation based Fault Localization}

MBFL is a fault localization technique that leverages mutation analysis to localize faulty code entities by analyzing the relationship between test outcomes and the kill information of mutants~\cite{moon2014ask,papadakis2015metallaxis}.

MBFL operates under the hypothesis that \textit{faulty code entities are more likely to distinguish between the original program and its defective variants (mutants)}. The core intuition is that the faulty code will cause the mutant's behavior to deviate from the original program, leading to differential test outcomes. This principle is formalized through suspiciousness formulas that quantify the correlation between mutants' execution outcomes and potential faulty code entities.

\subsubsection{Workflow of MBFL}


The workflow of MBFL is as follows:

\begin{enumerate}

    \item \textbf{Mutant Generation}: Generate a set of mutants $\mathcal{M}$ by applying mutation operators (e.g., arithmetic, conditional) to the original code.

    \item \textbf{Test Execution}: Execute each mutant $m \in \mathcal{M}$ with test cases $\mathcal{T}$, resulting in a binary matrix $\mathbf{K}$ where $K_{i,j} = 1$ if mutant $m_i$ is killed by test $t_j$, else $0$.

    \item \textbf{Suspiciousness Calculation}: Compute suspiciousness $Sus(m_i)$ for each mutant $m_i$. For each mutant $m_i$, we compute four statistical values: $a_{kf}(m_i)$ (killed by failing tests), $a_{kp}(m_i)$ (killed by passing tests), $a_{nf}(m_i)$ (not killed by failing tests), and $a_{np}(m_i)$ (not killed by passing tests). Common suspiciousness formulas include:

    \begin{table}[htb]
    \centering
    \caption{The Suspiciousness Formulas of MBFL}
    \label{tab:mbfl_formulas}
    \begin{tabular}{cl}
    \toprule
    \textbf{Formula} & \multicolumn{1}{c}{\textbf{Definition}} \\
    \hline
    Jaccard~\cite{chen2002pinpoint} & $Sus(m_i) = \frac{a_{kf}(m_i)}{a_{kf}(m_i)+a_{nf}(m_i)+a_{kp}(m_i)}$ \\
    \hline
    Tarantula~\cite{jones2005empirical} & $Sus(m_i) = \frac{\frac{a_{kf}(m_i)}{a_{kf}(m_i)+a_{nf}(m_i)}}{\frac{a_{kf}(m_i)}{a_{kf}(m_i)+a_{nf}(m_i)}+\frac{a_{kp}(m_i)}{a_{kp}(m_i)+a_{np}(m_i)}}$ \\
    \hline
    Ochiai~\cite{abreu2006evaluation} & $Sus(m_i) = \frac{a_{kf}(m_i)}{\sqrt{(a_{kf}(m_i)+a_{nf}(m_i))\times(a_{kf}(m_i)+a_{kp}(m_i))}}$ \\
    \hline
    Op2~\cite{naish2011model} & $Sus(m_i) = a_{kf}(m_i) - \frac{a_{kp}(m_i)}{a_{kp}(m_i)+a_{np}(m_i)+1}$ \\
    \hline
    Dstar~\cite{wong2013dstar} & $Sus(m_i) = \frac{a_{kf}^{*}(m_i)}{a_{kp}(m_i)+a_{nf}(m_i)}$ \\
    \hline
    GP13~\cite{yoo2012evolving} & $Sus(m_i) = a_{kf}(m_i) + \frac{a_{kf}(m_i)}{2\times a_{kp}(m_i)+a_{np}(m_i)}$ \\
    \hline
    \end{tabular}
    \end{table}

    The suspiciousness of a code entity $e$ is computed as the maximum suspiciousness of its mutants~\cite{papadakis2015metallaxis}.

    \item \textbf{Rank and Analysis}: Rank code entities $e$ by $Sus(e)$ to prioritize debugging efforts. 

\end{enumerate}

\subsubsection{Research Trends in MBFL}
\label{sec:mbfl-trends}



MBFL research focuses on two objectives: efficiency improvement and effectiveness enhancement, with optimization efforts targeting each step of the MBFL process. 
Early MBFL research primarily focused on suspiciousness calculation, aiming to achieve better fault localization performance through improved measurement of mutant suspiciousness (i.e., fault probability). Initial studies such as Metallaxis~\cite{papadakis2015metallaxis} and MUSE~\cite{moon2014ask} laid the foundation for this research direction. Recent advances have introduced more sophisticated approaches: entropy-based methods like SMBFL~\cite{bayati2020smbfl} prioritize code entities using information theory to enhance ranking accuracy; bias elimination techniques such as Delta4Ms~\cite{du2022improving,liu2024delta4ms} model and measure mutant bias to mitigate distortions in suspiciousness scores; dynamic weighting schemes like BLMu~\cite{wang2025systematic} assign different weights to mutants according to their kill information; and multi-fault approaches like HMBFL~\cite{li2021hmbfl} extend formulas by analyzing the many-to-many relationships between Higher-Order Mutants (HOMs) and code entities to improve MBFL effectiveness in multiple fault scenarios.
Following the initial establishment of MBFL research, researchers started investigating mutant generation and selection strategies to reduce computational overhead and improve fault localization effectiveness.
Early work by Papadakis et al.~\cite{papadakis2015metallaxis} and Liu et al.~\cite{liu2017statement} addressed computational costs through selective mutation techniques. With the advancement of higher-order mutation and neural mutation approaches, researchers have pursued effectiveness improvements through mutant-centric methodologies. Higher-order mutation techniques like HOTFUZ~\cite{jang2022hotfuz}, SFClu~\cite{wu2025boosting}, and SCOPE~\cite{liu2025scope} combine multiple mutations to create HOMs that better simulate real faults while reducing computational overhead. Neural-based approaches, such as Neural-MBFL~\cite{du2024neural}, leverage deep learning to generate semantically meaningful mutants that mimic diverse and realistic faults.
Kill information is the feature representation of mutation testing results, while the kill matrix serves as its structured data representation, constituting a basic component in MBFL. Early MBFL research demonstrated preliminary discussions of kill conditions in foundational studies by Papadakis et al.~\cite{papadakis2015metallaxis} and Moon et al.~\cite{moon2014ask}, while Li et al.~\cite{li2017transforming} provided systematic analysis of kill conditions. As research on kill information evolved, the concept of kill matrix gradually emerged~\cite{kim2021ahead,kim2022predictive}. Current research on kill matrix primarily focuses on prediction-based cost reduction approaches: SIMFL~\cite{kim2021ahead, kim2023learning} employs statistical inference on historical kill matrices to enable instant localization for new faults by decoupling expensive mutation analysis from debugging; Seshat~\cite{kim2022predictive} predicts entire kill matrices using deep learning on source code's natural language elements, avoiding execution overhead; and Xu et al.~\cite{xu2024predictive} propose a predictive kill matrix model utilizing execution results of a subset of mutants and their program features to forecast the remaining kill matrix, reducing the time overhead of mutation analysis.

Recent studies~\cite{jang2022hotfuz, kim2023learning, du2022improving, liu2024delta4ms, liu2025scope} have observed a subtle but critical issue that significantly impacts MBFL effectiveness: the presence of \emph{noise} in the fault localization process. These \emph{noise} phenomena~\cite{liu2025scope} manifest as false mutant-test relationships that mislead suspiciousness calculations and degrade localization accuracy. The complexity and diversity of noise sources make direct noise identification computationally prohibitive, as it requires extensive analysis of individual mutant behaviors and their semantic relationships with program faults. Existing approaches have attempted to address this challenge through different strategies, including HOM generation strategies to distribute couplings more evenly~\cite{jang2022hotfuz}, statistical coupling analysis to remove low-coupling mutants~\cite{kim2023learning}, and bias estimation techniques~\cite{du2022improving, liu2024delta4ms}. 
While some progress has been made, they primarily focus on mitigating noise effects rather than remove noise at its source. Most existing techniques operate by adjusting the final suspiciousness scores after noise has already influenced the calculations, rather than directly addressing the noise within the kill matrix itself, leaving room for techniques that address noise at the matrix level.

\textit{Research Scope.} 
To maintain a focused investigation and ensure the clarity of our findings, we establish clear boundaries for our research scope. This study concentrates on classical MBFL techniques, with particular emphasis on addressing noise suppression in the kill matrix. We restrict our analysis to traditional first-order mutants and conventional suspiciousness calculation formulas. Our investigation deliberately excludes techniques that leverage historical mutation information or predictive kill matrix modeling. We also set aside methods primarily aimed at computational cost reduction through test or mutant selection, as well as hybrid approaches that combine MBFL with spectrum-based or deep learning-based fault localization. This well-defined scope enables us to examine the fundamental problem of noise suppression in kill matrices under MBFL's basic assumption that mutants originating from faulty statements are more likely to be killed by failing tests than those from correct statements.

\subsection{Signal Processing and Its Application in Software Engineering}
\label{sec:signal-processing}

Signal processing is a foundational discipline for analyzing and modifying signals to extract meaningful information~\cite{proakis2007digital}. In this section, we first introduce core concepts relevant to our method and then discuss their applications in software engineering.

\subsubsection{Signal Processing Foundations}
\label{sec:signal-basics}

\paragraph{Fourier Transform and Filtering}
The \textit{Fourier Transform} decomposes a signal into its frequency components, enabling separation of \textit{low-frequency} (smooth, globally consistent patterns) and \textit{high-frequency} (random, local noise) components~\cite{oppenheim1999discrete, bracewell1978fourier}. For a 2D signal (e.g., the kill matrix $\mathbf{K}$), the transform is defined as:
\begin{equation}
    \hat{K}(u,v) = \sum_{x=0}^{M-1} \sum_{y=0}^{N-1} K(x,y) e^{-j2\pi(ux/M + vy/N)},
\end{equation}
where $\hat{K}(u,v)$ represents the frequency component at coordinates $(u,v)$~\cite{gonzalez2018digital}. 

\paragraph{Low-Pass Filtering}
A \textit{low-pass filter} suppresses high-frequency components (e.g., noise) while retaining low-frequency patterns (fault-relevant signals)~\cite{gonzalez2018digital, oppenheim1999discrete}. For a 2D matrix $\mathbf{K}$, this is applied as:
\begin{equation}
    \mathbf{K}_{\text{filtered}} = \mathcal{F}^{-1}\left[ H(u,v) \cdot \hat{K}(u,v) \right],
\end{equation}
where $H(u,v)$ is a filter mask that sets $H(u,v) = 1$ for low frequencies and $H(u,v) = 0$ for high frequencies (e.g., a Gaussian mask)~\cite{gonzalez2018digital}.

\paragraph{Signal-to-Noise Ratio (SNR)}
The \textit{Signal-to-Noise Ratio (SNR)} is a fundamental metric in signal processing that quantifies the level of desired signal relative to background noise~\cite{proakis2007digital}. A higher SNR indicates that the meaningful signal components are more prominent compared to noise, which is crucial for effective signal processing operations. In the context of denoising, improving the SNR before applying filtering techniques enhances the effectiveness of noise suppression algorithms, as they can better distinguish between signal and noise components~\cite{gonzalez2018digital}. 

\subsubsection{Applications in Software Engineering}
\label{sec:signal-applications}

Signal processing techniques have been applied to software engineering tasks across multiple domains, as follows:

\begin{itemize}

\item \textit{Defect Prediction:}
Yang et al.~\cite{yang2018software} applied Fourier transforms to Boolean functions of software metrics, achieving higher defect prediction accuracy than random forests. Takan~\cite{takan2019fourier} proposed Fourier-based test case generation for finite state machines (FSMs), reducing test suite size while improving fault detection ratios. 

\item \textit{Mutation Testing Enhancement:}
Takan and Ayav~\cite{takan2020mutant} leveraged Fourier analysis to prioritize critical mutants in FSM testing, reducing computational costs without sacrificing fault detection. Jammalamadaka and Parveen~\cite{jammalamadaka2022equivalent} combined wavelet transforms with rain optimization to detect equivalent mutants, achieving 85.17\% classification accuracy.



\item \textit{Code Clone Detection:} 
Karus and Kilgi~\cite{karus2015code} used wavelet analysis for multi-scale code comparison, achieving superior clone detection over conventional methods. Cao et al.~\cite{cao2020ftclnet} designed FTCLNet, a Fourier-based deep learning model for vulnerability detection that outperformed sequence-based methods on buffer overflow datasets.

\end{itemize}

These successful applications of signal processing across multiple software engineering domains provide valuable insights for our research and demonstrate their potential for addressing challenges in fault localization.


\paragraph{Connection to Our Method}
In MBFL, the kill matrix can be viewed as a signal matrix, making it feasible to apply signal processing methods. The kill matrix $\mathbf{K}$ inherently contains two types of information:
\begin{itemize}
    \item \textbf{Signal}: Low-frequency patterns caused by the actual fault (e.g., consistent kill status across related mutants and tests).
    \item \textbf{Noise}: High-frequency perturbations from incidental kills or uninformative mutants.
\end{itemize}
By applying 2D low-pass filtering (Equation 2), we can isolate the \textit{signal component} that correlates with true faults, directly addressing the noise problem in MBFL. 

\section{DKMR: Denoising based Kill Matrix Refinement}

\subsection{Signal View of Kill Matrix}

\textbf{Kill Matrix as Signal}:
MBFL relies on kill matrix to capture the execution behavior of mutants against test cases. When mutants and test cases are systematically ordered, the resulting kill matrix can be viewed as a two-dimensional signal that encodes meaningful fault-related information. 

In the mutant dimension (rows), mutants are ordered by their source code locations, which approximates a signal that reflects program behavior changes in both data flow and control flow. In the test dimension (columns), test cases are grouped by test suites and ordered alphabetically within each suite, which approximates a signal that represents impact variations when test inputs test similar functionality but target different code entities. This ordered structure transforms the kill matrix from a mere collection of binary values into a structured signal by creating structural locality (where mutants inserted at nearby code lines exhibit similar execution behaviors) and establishing coherence (where functionally related tests demonstrate consistent kill patterns), allowing meaningful fault-related patterns to manifest as coherent regions of consistent kill behaviors. For instance, mutants inserted near actual fault locations tend to exhibit similar kill patterns across failing test cases, which creates low-frequency signal components that indicate fault proximity. Conversely, mutants far from faults show more random kill behaviors, which contributes to higher-frequency noise components in the signal. This frequency separation is further supported by MBFL's assumption that mutants from faulty statements should exhibit significantly higher kill rates compared to those from correct statements, naturally leading to meaningful fault signals manifesting as low-frequency components while noise appears as high-frequency components.



\textbf{Noise Sources in Kill Matrix}:
The kill matrix, as the structured representation of mutant-test relationships, contains various forms of noise that can obscure the true fault-related patterns.
To better understand the noise phenomena in the kill matrix and their impact on fault localization, we conduct a preliminary analysis as follows.

Noise commonly arises when mutants of correct statements are unexpectedly killed by failing test cases. 
This occurs when a mutant introduces changes that coincidentally align with the test's execution path, causing the failing test to pass or exhibit different error behaviors even though the mutated statement is not related to the actual fault. 
These noise sources can create false positive signals that mislead the suspiciousness calculation and degrade the fault localization accuracy.
Additionally, inadequate test cases or weak test oracles may lead to testing inadequacy that fails to distinguish between subtle behavioral differences, leading to inconsistent kill patterns where similar mutants exhibit different kill behaviors under similar test conditions.

Conversely, noise also manifests when mutants of faulty statements are not killed by failing test cases that should detect them. 
This happens when the mutation does not sufficiently alter the program's behavior to be detected by the available test cases, or when the test cases lack adequate coverage or sensitivity to reveal the mutant's impact. 
Furthermore, complex program semantics and intricate control flow dependencies can create situations where fault-related mutants survive unexpectedly, generating false negative signals that reduce the fault localization accuracy.

These noise sources are pervasive across different programs and test suites, creating a fundamental challenge in MBFL: distinguishing meaningful fault-related patterns from various forms of noise that naturally occur in kill matrix. 
The prevalence of such noise necessitates systematic approaches to enhance signal quality and suppress noise.

\subsection{DKMR Method}
\label{sec:dkmr}

To address the noise challenges in kill matrix, we propose DKMR (Denoising based Kill Matrix Refinement) that treats kill matrix as signals and applies signal processing techniques to enhance fault localization effectiveness.

DKMR consists of two main stages: (1) Signal enhancement that improves the SNR by combining weak kill and strong kill information to provide higher quality input for the subsequent denoising stage, and (2) Signal denoising that applies frequency domain filtering techniques to suppress high-frequency noise while preserving meaningful fault-related signals. 

\textit{Why DKMR Works.} 
The effectiveness of DKMR stems from the signal nature of kill matrix. Meaningful fault-related patterns manifest as consistent kill behaviors across related test cases, representing low-frequency signals that indicate proximity to faults. In contrast, high-frequency noise like false kills obscures true fault-related patterns. DKMR systematically enhances meaningful patterns while suppressing noise components by leveraging two complementary strategies: (1) signal enhancement that incorporates domain-specific knowledge from fault localization by using strong kills to reinforce weak kills, thereby improving the SNR, and (2) frequency domain filtering that applies efficient and general-purpose denoising techniques from signal processing to suppress high-frequency noise while preserving low-frequency fault-related patterns. 

\subsubsection{Signal Enhancement}

The signal enhancement process improves the SNR by systematically combining different types of kill information, which is crucial for the subsequent denoising stage. We begin by constructing two fundamental kill matrix based on mutant execution behavior.

For a mutant $m_i$ executed against test case $t_j$, we define kill information hierarchically:
\begin{itemize}
    \item \textbf{Weak Kill (W-Kill)}: Occurs when a mutant exhibits any detectable difference from the original program, including both outcome differences and internal execution detail differences (e.g., different exception types or error messages):
    \[
        \textit{W-Kill}(m_i, t_j) = 
        \begin{cases} 
        1 & \text{if } \text{result}(m_i, t_j) \neq \text{result}(P, t_j) \text{ OR } \\
        & \text{err\_msg}(m_i, t_j) \neq \text{err\_msg}(P, t_j), \\
        0 & \text{otherwise}.
        \end{cases}
    \]
    
    \item \textbf{Strong Kill (S-Kill)}: A stricter form of weak kill that occurs only when the mutant's test outcome differs from the original program's outcome, obtained by strengthening the weak kill condition:
    \[
        \textit{S-Kill}(m_i, t_j) = 
        \begin{cases} 
        1 & \text{if } \text{result}(m_i, t_j) \neq \text{result}(P, t_j), \\
        0 & \text{otherwise}.
        \end{cases}
    \]
\end{itemize}


Based on existing research~\cite{liu2024delta4ms, li2017transforming} demonstrating that weak kill information provides better MBFL effectiveness, we use the weak kill matrix as our foundation. However, recognizing that strong kill information contains less noise~\cite{li2017transforming}, we enhance the SNR to facilitate more effective denoising by incorporating strong kill information into the weak kill matrix:

\[
    M^E_{i,j} = 
    \begin{cases} 
    2 & \text{if } \textit{S-Kill}(m_i, t_j) = 1, \\
    1 & \text{if } \textit{W-Kill}(m_i, t_j) = 1, \\
    0 & \text{otherwise}.
    \end{cases}
\]

where $2$ denotes strong kill, $1$ denotes weak kill, and $0$ denotes survival. This enhancement strategy leverages the complementary strengths of both kill types: the comprehensive coverage of weak kill and the high confidence of strong kill.

\textbf{Matrix Ordering for Signal Enhancement}:
We make a simplifying assumption that systematic ordering of matrix dimensions can better satisfy the conditions required for effective signal processing. Based on this assumption, we apply the following ordering strategy:
\begin{itemize}
    \item \textit{Mutant Ordering (Rows)}: Mutants are sorted by their source code line numbers to preserve structural locality and group related mutants together.
    \item \textit{Test Ordering (Columns)}: Test cases from the same test suite are grouped together, and within each suite, test cases are sorted alphabetically (which implicitly reflects functional ordering) to enhance temporal coherence and group semantically related tests.
\end{itemize}

This ordering approach is reasonable for several reasons: (1) mutants from nearby code lines often exhibit similar behavioral patterns due to code locality, making row-wise ordering beneficial for pattern recognition; (2) test cases from the same test suite typically target related functionalities, and alphabetical ordering within suites often reflects functional progression, making this grouping and ordering effective for signal coherence; and (3) the resulting structured arrangement creates a more regular signal pattern that is amenable to frequency-domain analysis.

\subsubsection{Signal Denoising}
The signal denoising process applies frequency domain filtering to suppress high-frequency noise while preserving meaningful low-frequency patterns. 

\textbf{Frequency Domain Analysis}: 
The enhanced kill matrix $M^E \in \{0, 1, 2\}^{N \times M}$ is transformed into the frequency domain using the 2D Fourier Transform:
\[
F(u, v) = \sum_{x=1}^{N} \sum_{y=1}^{M} M^E_{x,y} e^{-j2\pi\left(\frac{ux}{N} + \frac{vy}{M}\right)}
\]

This transformation enables the separation of meaningful fault-related patterns (low frequencies) from noise and random variations (high frequencies). 

\textbf{Low-pass Filtering for Noise Suppression}: We then apply a low-pass filter with cutoff frequency $D_0$ to suppress high-frequency noise:
\[
H(u, v) = 
\begin{cases} 
1 & \text{if } \sqrt{\left(\frac{u}{N}\right)^2 + \left(\frac{v}{M}\right)^2} \leq D_0 \\
0 & \text{otherwise}
\end{cases}
\]
The threshold $D_0$ is empirically set to a fixed value (e.g., $D_0=0.3$) to achieve optimal balance between noise suppression and pattern preservation. This parameter choice is validated through our preliminary experiments.

\textbf{Matrix Reconstruction}: 
The refined matrix $M^R$ is reconstructed via inverse Fourier transform:
\[
M^R_{x,y} = \frac{1}{NM} \sum_{u=1}^{N} \sum_{v=1}^{M} F(u, v)H(u, v) e^{j2\pi\left(\frac{ux}{N} + \frac{vy}{M}\right)}
\]
This reconstruction process produces a refined kill matrix that preserves meaningful fault-related patterns while suppressing noise.

\textbf{Matrix Normalization}:
Finally, we apply max-min normalization to scale all values into the $[0, 1]$ range:
\[
M'_{x,y} = \frac{M^R_{x,y} - \min(M^R)}{\max(M^R) - \min(M^R)}
\]

The normalized matrix $M'$ represents the final refined kill matrix with enhanced SNR used for subsequent fault localization calculations.

\section{MBFL-DKMR: Comprehensive Fault Localization Framework}
\label{sec:mbfl-dkmr}

\subsection{Overview and Workflow}

\begin{figure}[htb]
\centering
\includegraphics[scale = 0.33]{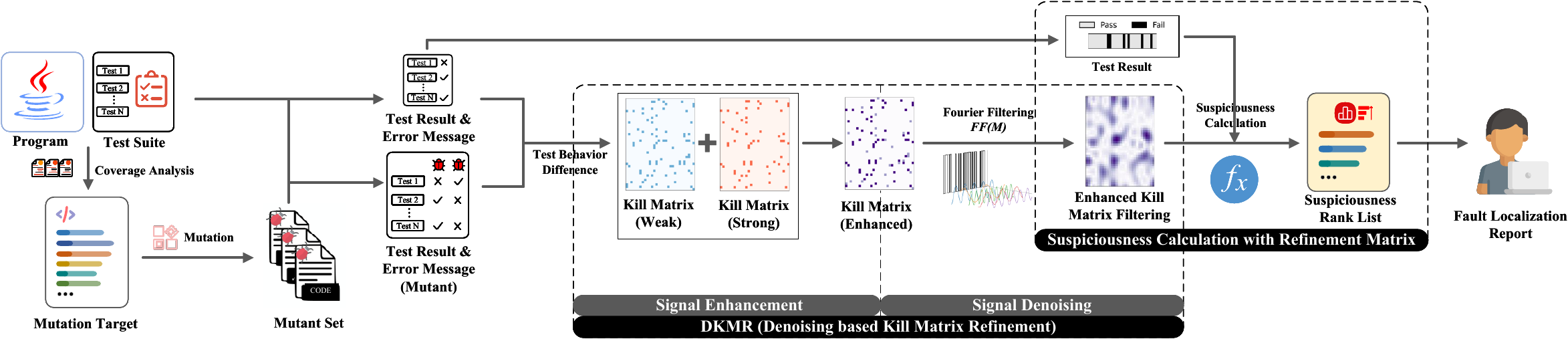}
\caption{Workflow of MBFL-DKMR}
\label{fig:mbfl-dkmr}
\end{figure}

To provide a complete fault localization solution, we propose MBFL-DKMR, a comprehensive framework that integrates the DKMR method with suspiciousness calculation to achieve effective fault localization. MBFL-DKMR consists of two main components: (1) DKMR component that systematically enhances kill matrices through signal enhancement and denoising techniques, (2) Suspiciousness calculation with refined matrices that computes fault localization scores using fuzzy values. The overall workflow of DKMR is illustrated in Figure~\ref{fig:mbfl-dkmr}.

The DKMR component first improves the SNR by combining weak kill and strong kill information, then applies frequency domain filtering techniques to denoise the kill matrix while preserving meaningful fault-related signals. The suspiciousness calculation component computes suspiciousness scores using the refined matrices with continuous values ranging from 0 to 1, generating more accurate suspiciousness scores for each program element. 
The DKMR method has been detailed in the Section~\ref{sec:dkmr}, in the following subsection, we focus on introduce the suspiciousness calculation component, which leverages the refined matrix to compute suspiciousness scores for effective fault localization.

\subsection{Suspiciousness Calculation with Refined Matrix}

The suspiciousness calculation leverages the refined matrix $M'$ with fuzzy values ranging from 0 to 1. Unlike classical MBFL approaches that use boolean kill matrix, our approach enables fuzzy suspiciousness calculation that provides more nuanced information for fault localization.

\textbf{Fuzzy Kill Statistics}:
For each mutant $m_i$, we compute four kill statistics using fuzzy matrix values:
\[
\begin{aligned}
a'_{kf}(m_i) &= \sum_{j=1}^M M'_{i,j} \cdot I_{f_j} \\
a'_{kp}(m_i) &= \sum_{j=1}^M M'_{i,j} \cdot (1 - I_{f_j}) \\
a'_{nf}(m_i) &= \sum_{j=1}^M (1 - M'_{i,j}) \cdot I_{f_j} \\
a'_{np}(m_i) &= \sum_{j=1}^M (1 - M'_{i,j}) \cdot (1 - I_{f_j})
\end{aligned}
\]

where $I_{f_j}$ indicates whether test case $t_j$ fails (1) or passes (0). These statistics quantify the accumulated behavior of each mutant across test outcomes, based on the nuanced relationships captured by the refined fuzzy matrix.

\textbf{Mutant-level Suspiciousness}:
The mutant-level suspiciousness is calculated using standard MBFL metrics (e.g., Ochiai, DStar, Jaccard) applied to the fuzzy statistics:
\[
\textit{Sus}(m_i) = f(a'_{kf}(m_i), a'_{kp}(m_i), a'_{nf}(m_i), a'_{np}(m_i))
\]

For example, using the Ochiai metric:
\[
\textit{Sus}_{\text{Ochiai}}(m_i) = \frac{a'_{kf}(m_i)}{\sqrt{(a'_{kf}(m_i)+a'_{kp}(m_i))(a'_{kf}(m_i)+a'_{nf}(m_i))}}
\]

\textbf{Statement-level Localization}:
To identify the most suspicious statement $s$, we aggregate mutant-level scores using maximum aggregation:
\[
\textit{Sus}(s) = \max_{m_i \in \mathcal{M}_s} \textit{Sus}(m_i)
\]

This approach focuses on the strongest evidence from mutants associated with statement $s$, prioritizing the most likely fault location based on the refined matrix analysis.

\section{Experimental Design}\label{sec:ExperimentalDesign}

\subsection{Research Questions}

\textbf{RQ1:} How effective is the DKMR component in suppressing noise?

RQ1 investigates the effectiveness of our DKMR component in noise suppression and fault localization accuracy improvement. We conduct an ablation study comparing three MBFL-DKMR variants to demonstrate the progressive impact of signal denoising and signal enhancement within the DKMR component.

\textbf{RQ2:} How does MBFL-DKMR compare with state-of-the-art MBFL techniques?

RQ2 evaluates the performance of MBFL-DKMR against advanced MBFL approaches, specifically BLMu and Delta4Ms. This research question aims to validate whether our signal filtering approach can achieve competitive or better results compared to existing state-of-the-art MBFL techniques.

\textbf{RQ3:} What is the time cost of MBFL-DKMR compared to existing MBFL techniques?

RQ3 evaluates the computational efficiency of MBFL-DKMR by comparing its execution time with existing MBFL techniques. We analyze the time cost across three key stages: (1) mutation analysis, which includes mutant generation and test execution, (2) DKMR refinement, which consists of signal enhancement and signal denoising, and (3) suspiciousness calculation. The objective is to assess whether the additional computational cost introduced by the DKMR process is worthwhile compared to the improvements of fault localization effectiveness.

\begin{table}[htb]
    \centering
    \caption{Statistics of subject programs}
    \label{tab:dataset}%
    \scalebox{1}{
    \begin{tabular}{ccrrr}
    \toprule
    \textbf{Project} & \textbf{Description} & \multicolumn{1}{c}{\textbf{\#Version}} & \multicolumn{1}{c}{\textbf{Avg. \#TestCase}} & \multicolumn{1}{c}{\textbf{Avg. \#kLOC}} \\
    \midrule
    \multicolumn{1}{l}{Chart} & \multicolumn{1}{l}{JFreeChart} & 26    & 1,814  & 203  \\
    \multicolumn{1}{l}{Closure} & \multicolumn{1}{l}{Google Closure compiler} & 174   & 7,027  & 139  \\
    \multicolumn{1}{l}{Lang} & \multicolumn{1}{l}{Apache commons-lang} & 64    & 1,815  & 52  \\
    \multicolumn{1}{l}{Math} & \multicolumn{1}{l}{Apache commons-math} & 106   & 2,513  & 116  \\
    \multicolumn{1}{l}{Mockito} & \multicolumn{1}{l}{Mockito framework} & 38    & 1,140  & 19  \\
    \multicolumn{1}{l}{Time} & \multicolumn{1}{l}{Joda-Time} & 26    & 3,918  & 68  \\
    \multicolumn{1}{l}{Cli} & \multicolumn{1}{l}{Commons-cli} & 39    & 262   & 6  \\
    \multicolumn{1}{l}{Codec} & \multicolumn{1}{l}{Commons-codec} & 18    & 440   & 11  \\
    \multicolumn{1}{l}{Collections} & \multicolumn{1}{l}{Commons-collections} & 4     & 15,583  & 67  \\
    \multicolumn{1}{l}{Compress} & \multicolumn{1}{l}{Commons-compress} & 47    & 421   & 31  \\
    \multicolumn{1}{l}{Csv} & \multicolumn{1}{l}{Commons-csv} & 16    & 198   & 3  \\
    \multicolumn{1}{l}{Gson} & \multicolumn{1}{l}{Gson} & 18    & 988   & 14  \\
    \multicolumn{1}{l}{JacksonCore} & \multicolumn{1}{l}{Jackson-core} & 26    & 356   & 34  \\
    \multicolumn{1}{l}{JacksonDatabind} & \multicolumn{1}{l}{Jackson-databind} & 112   & 1,622  & 96  \\
    \multicolumn{1}{l}{JacksonXml} & \multicolumn{1}{l}{Jackson-dataformat-xml} & 6     & 152   & 8  \\
    \multicolumn{1}{l}{Jsoup} & \multicolumn{1}{l}{Jsoup} & 93    & 454   & 15  \\
    \multicolumn{1}{l}{JxPath} & \multicolumn{1}{l}{Commons-jxpath} & 22    & 347   & 29  \\
    \midrule
    \multicolumn{2}{c}{\textbf{Summary}} & 835   & 2,587  & 76  \\
    \bottomrule
    \end{tabular}%
    }
\end{table}%

\subsection{Dataset}

The Defects4J benchmark~\cite{just2014defects4j} is a valuable collection of repeatable faults found in major open-source applications, designed to support software engineering research. In our experiment, we utilized the Defects4J v2.0.0 benchmark~\cite{gay2020defects4j}, which is widely used in software engineering research, particularly in areas such as software testing, fault localization, automated program repair (APR), and regression testing selection~\cite{bennett2022some}. Table~\ref{tab:dataset} provides a detailed overview of the subject programs drawn from the Defects4J v2.0.0 benchmark. It includes a range of open-source projects, each varying in complexity, size, and the number of test cases. The table lists statistics such as the number of versions studied (\#Version), average number of test cases (Avg. \#TestCase), and average number of thousand lines of code (Avg. \#kLOC), which gives an indication of the project's size. This diversity in projects helps in assessing the robustness and generalizability of our fault localization approach across different real-world software applications. Collectively, the projects have a total of 835 versions, with an average of approximately 2,587 test cases and 76 kLOC.

\subsection{Evaluation Metric}

We adopt the following metrics to evaluate fault localization effectiveness:

\begin{enumerate}

    \item \emph{TOP-N} counts the number of faults localized within top n ranks~\cite{zou2019empirical}.
    
    According to previous studies, 73.58\% of developers only inspect \emph{TOP-5} code entities~\cite{li2019deepfl}, and we also adopt \emph{TOP-N} ($N \in \{1,3,5\}$).

    \item \emph{MAP}~\cite{moffat2008rank} (Mean Average Precision) measures the mean of the Average Precision (AP) values for all faults in a group. AP (Average precision) can be calculated as follows:

    \begin{equation}
    AP = \sum_{i=1}^M \frac{P(i) \times pos(i)}{Number\ of\ faulty\ statements}
    \end{equation}
    
    $pos(i)$ is a Boolean function, where $pos(i) = 1$ indicates the $i$-th statement is faulty, otherwise $pos(i) = 0$. $P(i)$ is the precision of localization at each rank $i$.

    \item \emph{EXAM}~\cite{zou2019empirical,liu2018optimal} (Examination Score) quantifies the effort required to locate the fault, expressed as the proportion of examined statements before reaching the fault:
    
    \begin{equation}
     \textit{EXAM} = \frac{\textit{rank}}{N},
    \label{eq:EXAM}
    \end{equation}
    
    Here, $\textit{rank}$ denotes the position of the faulty statement in a list sorted by descending suspiciousness, and $N$ is the total number of statements. The rank is computed as:
    
    \begin{equation}
     rank = \frac{(m+1) + (m+n)}{2},
     \label{eq:rank}
    \end{equation}
    
    where $m$ is the number of non-faulty statements ranked higher than the fault, and $n$ is the count of statements sharing the same suspiciousness score as the fault.

    \item \emph{Wilcoxon signed-rank test}~\cite{wilcoxon1945individual, liu2024delta4ms} is a non-parametric statistical test used to compare two related samples and determine if their medians are significantly different.
    We employ this test to statistically evaluate whether two fault localization techniques have significantly different effectiveness.
    Specifically, we formulate the null and alternative hypotheses as follows: $H_0: median_1 = median_2$ and $H_1: median_{1} \neq median_{2}$, where $median_{1}$ and $median_{2}$ denote the median \emph{EXAM} for the two fault localization methods, respectively. 
    In this paper, we set the significance level to 0.05 following previous studies~\cite{pearson2017evaluating, lei2022feature}. If the p-value for the \emph{Wilcoxon signed-rank test} is lower than this threshold, the null hypothesis is rejected, indicating that the median EXAM scores are significantly different between the two methods. 
    The \emph{Wilcoxon signed-rank test} can be conducted in three forms: 
    (1) \emph{two-sided test} determines whether there is a significant difference in median EXAM scores regardless of direction; 
    (2) \emph{one-sided (less) test} determines whether one method has significantly lower median EXAM scores than the other; 
    and (3) \emph{one-sided (greater) test} determines the opposite.
    
    \item \emph{Cliff's Delta} effect size~\cite{macbeth2011cliff, amario2022understanding} quantifies the magnitude and direction of differences between two methods by measuring the proportion of data range accounted for by their difference. 
    We adopt widely used benchmarks for interpreting \emph{Cliff's Delta} magnitude: $|\delta| \approx 0.147$ (small effect), $|\delta| \approx 0.33$ (medium effect), and $|\delta| \approx 0.474$ (large effect)~\cite{meissel2024using}.    
    By combining the \emph{Wilcoxon signed-rank test} and \emph{Cliff's Delta} effect size, we can comprehensively evaluate whether the differences between fault localization methods are both statistically significant and practically meaningful.

\end{enumerate}

\subsection{Experimental Configuration}

In our experiments, we employ six widely-used suspiciousness formulas for MBFL: Dstar~\cite{wong2013dstar}, GP13~\cite{yoo2012evolving}, Jaccard~\cite{chen2002pinpoint}, Ochiai~\cite{abreu2006evaluation}, Op2~\cite{naish2011model}, and Tarantula~\cite{jones2005empirical}. We utilize the Major mutation testing tool with its default mutation operators to generate mutants for MBFL implementation. The fault localization process employs the test suites provided by the Defects4J dataset. All experiments were conducted on Ubuntu 20.04.6 LTS with an Intel(R) Core(TM) i9-13900K CPU (3.00GHz base frequency, 24 cores, 32 threads) and 128GB RAM.

\section{Results Analysis}\label{sec:ResultsAnalysis}

\subsection{RQ1: How effective is the DKMR in suppressing noise?}
\label{subsec:Results4RQ1}

In RQ1, we evaluate the effectiveness of our DKMR component in suppressing noise and improving fault localization accuracy. Specifically, we investigate how the signal denoising and signal enhancement in DKMR contribute to improved fault localization performance. For this evaluation, we conduct an ablation study comparing three variants: MBFL-DKMR (complete method with both signal enhancement and signal denoising), MBFL-DKMR$_{W}$ (DKMR with signal denoising only, without signal enhancement), and Metallaxis (traditional MBFL technique without DKMR). The effectiveness of noise suppression is validated through the improvement in fault localization performance across these variants. We employ Top-N, MAP, and EXAM to evaluate the fault localization effectiveness. Additionally, we statistically validate the significance differences between the three MBFL-DKMR variants using the Wilcoxon signed-rank test with a significance level of 0.05 and calculate Cliff's Delta effect size to quantify the magnitude of differences between techniques.

\begin{table}[htb]
  \centering
  \caption{Effectiveness Evaluation of MBFL-DKMR in terms of Top-N and MAP}
    \begin{tabular}{lrrrr}
    \toprule
    \multicolumn{1}{c}{\textbf{Method}} & \multicolumn{1}{c}{\textbf{Top-1}} & \multicolumn{1}{c}{\textbf{Top-3}} & \multicolumn{1}{c}{\textbf{Top-5}} & \multicolumn{1}{c}{\textbf{MAP}} \\
    \midrule

    MBFL-DKMR & \boldmath{}\textbf{129$_{G}$}\unboldmath{} & \boldmath{}\textbf{282$_{O}$}\unboldmath{} & \boldmath{}\textbf{377$_{O}$}\unboldmath{} & \boldmath{}\textbf{0.2232$_{O}$}\unboldmath{} \\
MBFL-DKMR$_{W}$ & 98$_{J}$ & 265$_{J}$ & 370$_{J}$ & 0.2156$_{J}$ \\
    Metallaxis & 76$_{O}$ & 258$_{O}$ & 366$_{J}$ & 0.2074$_{O}$ \\
    \bottomrule
    \end{tabular}
  \label{tab:rq1-topn-map}

  \footnotesize{\textit{Note: Subscripts indicate the suspiciousness formula achieving optimal performance: \\ G (GP13), O (Ochiai), J (Jaccard), P (Op2), D (Dstar), and T (Tarantula).}}
\end{table}

Table~\ref{tab:rq1-topn-map} presents the Top-N and MAP results comparing MBFL-DKMR with its variant MBFL-DKMR$_{W}$ (signal denoising only) and the traditional Metallaxis technique (without DKMR). The results clearly demonstrate the effectiveness of the DKMR in suppressing noise and improving fault localization accuracy through progressive improvements.
The comparison between MBFL-DKMR$_{W}$ and Metallaxis demonstrates the effectiveness of the signal denoising stage. MBFL-DKMR$_{W}$ achieves 98 faults at Top-1 compared to Metallaxis's 76 faults, representing an improvement of 22 faults (28.9\% relative improvement). This improvement directly validates that the signal denoising stage of DKMR effectively suppresses high-frequency noise caused by coincidental kills and uninformative mutants, leading to more accurate suspiciousness calculations.
The comparison between MBFL-DKMR and MBFL-DKMR$_{W}$ demonstrates the additional benefit of signal enhancement. MBFL-DKMR achieves 129 faults at Top-1 compared to MBFL-DKMR$_{W}$'s 98 faults, representing an improvement of 31 faults (31.6\% relative improvement). This substantial improvement demonstrates that signal enhancement, which combines weak kill and strong kill information to improve SNR, makes the signal denoising stage more effective by providing higher quality input signals.
The MAP results further confirm these findings, with MBFL-DKMR achieving 0.2232, MBFL-DKMR$_{W}$ achieving 0.2156, and Metallaxis achieving 0.2074. The progressive improvements (3.5\% from MBFL-DKMR$_{W}$ to MBFL-DKMR, and 4.0\% from Metallaxis to MBFL-DKMR$_{W}$) validate that higher SNR enable more effective noise suppression, ultimately leading to improved fault localization performance.

\begin{figure}[htb]
\centering
\includegraphics[width=0.5\textwidth]{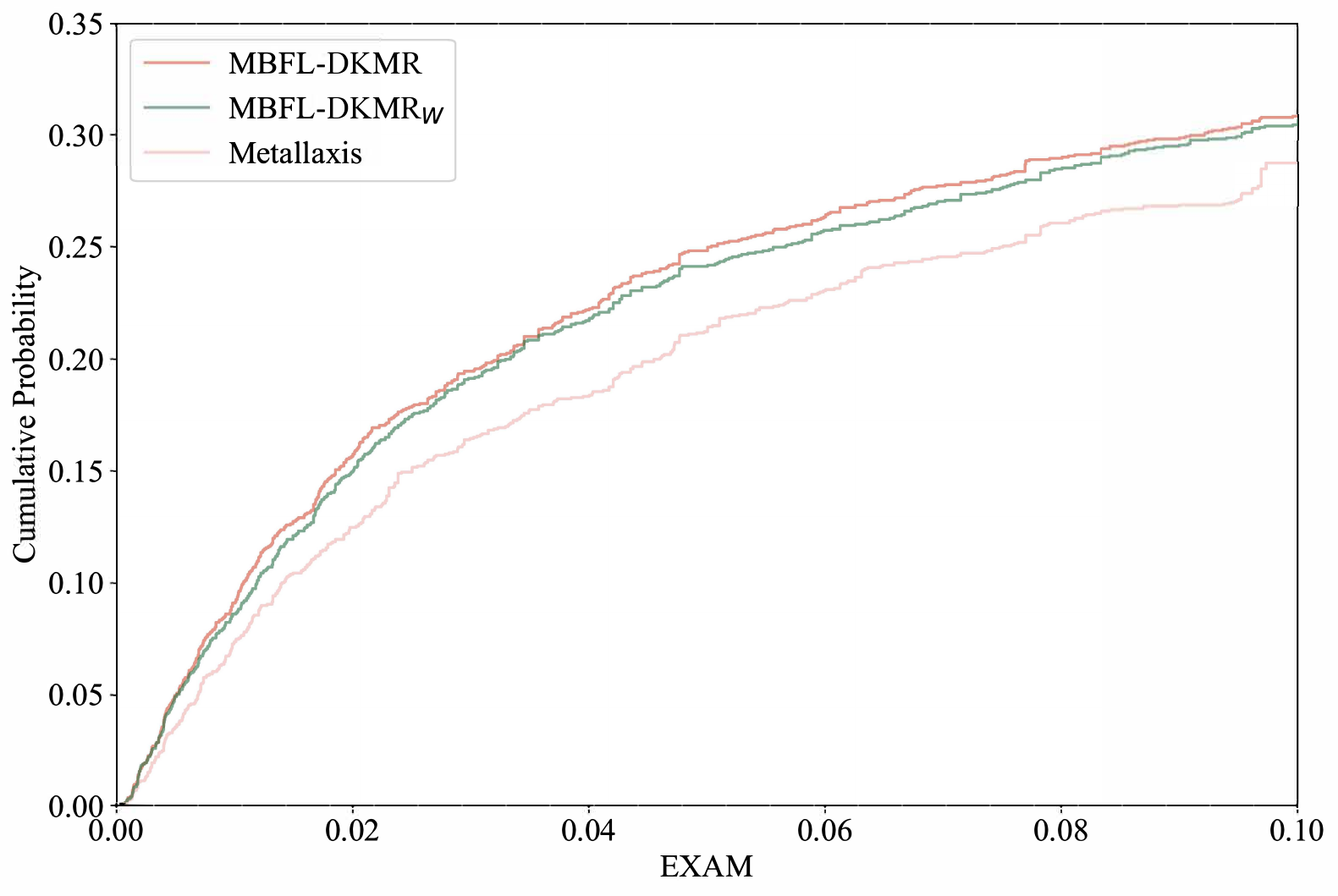}
\caption{Effectiveness Evaluation of MBFL-DKMR in terms of EXAM distribution}
\label{fig:rq1-exam}
\end{figure}

Fig.~\ref{fig:rq1-exam} illustrates the cumulative distribution of EXAM scores. The distributions demonstrate that MBFL-DKMR consistently achieves lower examination effort compared to MBFL-DKMR$_{W}$ and Metallaxis, indicating better performance. Specifically, MBFL-DKMR requires less examination effort to locate the same percentage of faults, thereby validating the effectiveness of the DKMR component in noise suppression and fault localization accuracy improvement.

\begin{table}[htb]
  \centering
  \caption{Statistical Significance Analysis of DKMR Effects}
    \begin{tabular}{cllllc}
    \toprule
    \multicolumn{2}{c}{\multirow{2}[4]{*}{\textbf{Comparison}}} & \multicolumn{3}{c}{\textbf{p-value}} & \multirow{2}[4]{*}{\tabincell{c}{\textbf{Cliff's}\\
        \textbf{Delta}}} \\
\cmidrule{3-5}          &       & \multicolumn{1}{c}{\textbf{two-sided}} & \multicolumn{1}{c}{\tabincell{c}{\textbf{one-sided}\\
        \textbf{(less)}}} & \multicolumn{1}{c}{\tabincell{c}{\textbf{one-sided}\\
        \textbf{(greater)}}} &  \\
    \midrule

    \multirow{2}[0]{*}{MBFL-DKMR v.s.} & \quad MBFL-DKMR$_{W}$ & 1.23E-45 & 6.15E-46 & 1.00E+00 & -0.0892 \\
          & \quad Metallaxis & 2.51E-121 & 1.25E-121 & 1.00E+00 & -0.1271 \\
    \bottomrule
    \end{tabular}
  \label{tab:rq1-significance}
\end{table}

To rigorously validate the effectiveness of the DKMR, we conducted statistical significance testing using the Wilcoxon signed-rank test on the EXAM scores. The results are presented in Table~\ref{tab:rq1-significance}. The statistical analysis confirms that MBFL-DKMR's performance improvements over both MBFL-DKMR$_{W}$ and Metallaxis are statistically significant (p-value $<$ 0.05). The negative Cliff's Delta values indicate that MBFL-DKMR consistently achieves lower EXAM scores (better performance) compared to both variants. The effect size is small to medium for MBFL-DKMR$_{W}$ (-0.0892) and medium for Metallaxis (-0.1271), demonstrating that both the signal denoising stage and the combined signal enhancement and denoising stages provide meaningful practical improvements.

In response to RQ1, our results demonstrate that the DKMR effectively suppresses noise and improves the effectiveness of fault localization. The comparison between MBFL-DKMR$_{W}$ and Metallaxis confirms that signal denoising alone successfully mitigates high-frequency noise. Furthermore, the comparison between MBFL-DKMR and MBFL-DKMR$_{W}$ validates that signal enhancement improves SNR through combining weak kill and strong kill information, making the signal denoising stage more effective. These progressive improvements validate that the kill matrix contains both valuable fault-related signals and high-frequency noise, and that DKMR through signal enhancement and denoising improves fault localization performance.

\subsection{RQ2: How does MBFL-DKMR compare with state-of-the-art MBFL techniques?}
\label{subsec:Results4RQ2}

RQ2 evaluates MBFL-DKMR's effectiveness by comparing it against state-of-the-art MBFL techniques. For this evaluation, we compare MBFL-DKMR against two advanced MBFL techniques: BLMu~\cite{wang2025systematic} (assigns different weights to mutants based on their contributions to fault localization by considering different kill types) and Delta4Ms~\cite{du2022improving, liu2024delta4ms} (models and eliminates mutant bias in suspiciousness calculation). The evaluation is performed on the complete Defects4J v2.0.0 benchmark, and we employ the same evaluation metrics as in RQ1: Top-N, MAP, and EXAM scores, along with statistical significance testing.

\begin{table}[htb]
  \centering
  \caption{Comparison between MBFL-DKMR and state-of-the-art MBFL Techniques in terms of Top-N and MAP}
    \begin{tabular}{lrrrr}
    \toprule
    \multicolumn{1}{c}{\textbf{Method}} & \multicolumn{1}{c}{\textbf{Top-1}} & \multicolumn{1}{c}{\textbf{Top-3}} & \multicolumn{1}{c}{\textbf{Top-5}} & \multicolumn{1}{c}{\textbf{MAP}} \\
    \midrule

    MBFL-DKMR & \boldmath{}\textbf{129$_{G}$}\unboldmath{} & \boldmath{}\textbf{282$_{O}$}\unboldmath{} & \boldmath{}\textbf{377$_{O}$}\unboldmath{} & \boldmath{}\textbf{0.2232$_{O}$}\unboldmath{} \\
    Delta4Ms & 103$_{J}$ & 268$_{J}$ & \boldmath{}\textbf{377$_{J}$} & 0.2140$_{J}$ \\
    BLMu  & 85$\ \ $ & 250$\ \ $ & 358$\ \ $ & 0.2144$\ \ $ \\
    \bottomrule
    \end{tabular}%
  \label{tab:rq2-topn-map}%

  \footnotesize{\textit{Note: Subscripts indicate the suspiciousness formula achieving optimal performance: \\ G (GP13), O (Ochiai), J (Jaccard), P (Op2), D (Dstar), and T (Tarantula).}}
\end{table}%

Table~\ref{tab:rq2-topn-map} presents the Top-N and MAP results comparing MBFL-DKMR against the state-of-the-art MBFL techniques BLMu and Delta4Ms on the complete Defects4J benchmark. The results demonstrate that MBFL-DKMR achieves competitive performance against these advanced techniques.
MBFL-DKMR successfully localizes 129 faults at Top-1, representing improvements of 44 faults over BLMu (85 faults) and 26 faults over Delta4Ms (103 faults). For Top-3 and Top-5 metrics, MBFL-DKMR achieves 282 and 377 faults respectively, compared to BLMu's 250 and 358, and Delta4Ms' 268 and 377. The MAP results further confirm MBFL-DKMR's effectiveness, with a value of 0.2232 compared to 0.2144 for BLMu and 0.2140 for Delta4Ms, representing relative improvements of 4.1\% and 4.3\% respectively. These results demonstrate that MBFL-DKMR outperforms the state-of-the-art MBFL techniques, validating its effectiveness.

\begin{figure}[htb]
\centering
\includegraphics[width=0.5\textwidth]{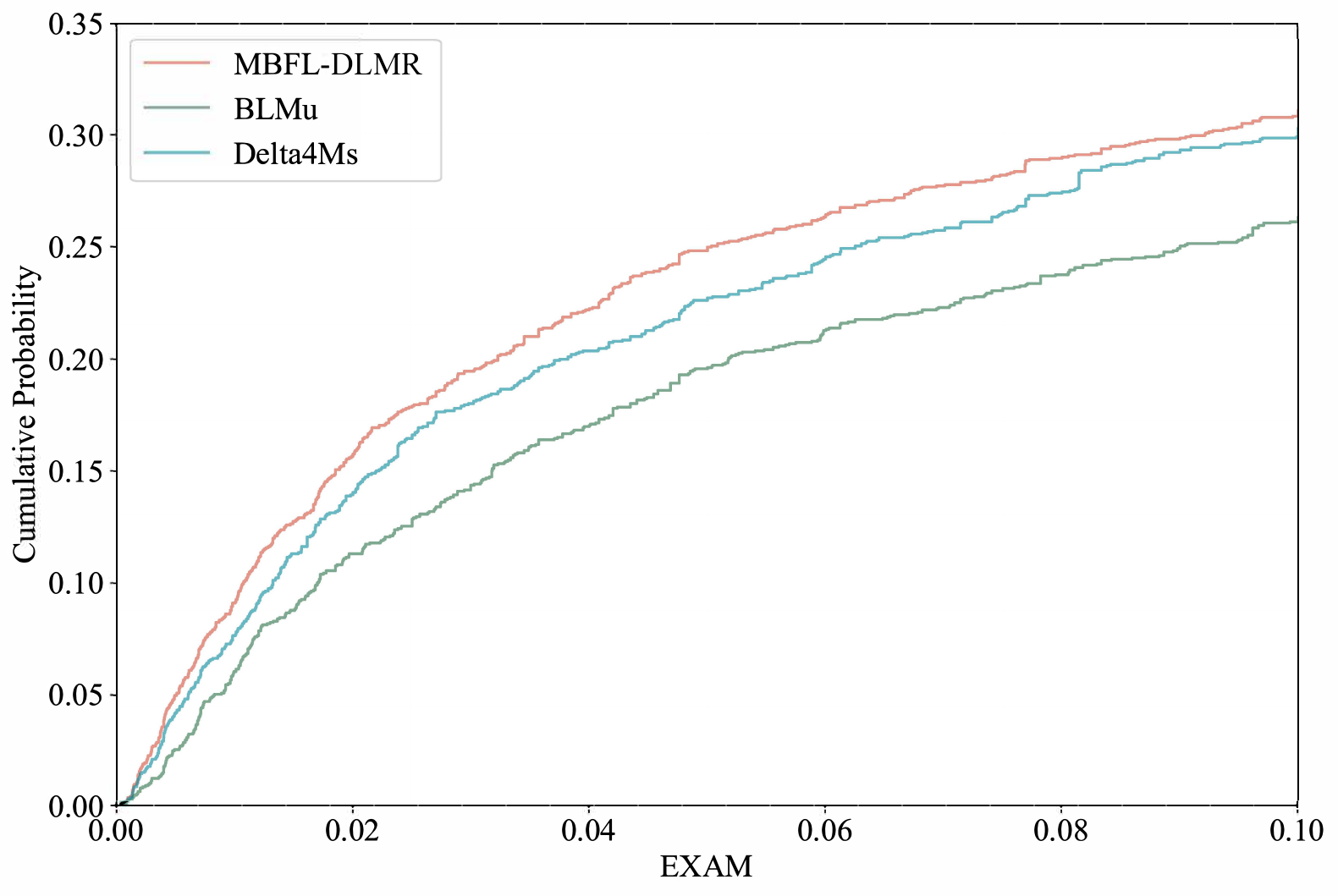}
\caption{Comparison between MBFL-DKMR and state-of-the-art MBFL Techniques in terms of EXAM distribution}
\label{fig:rq2-exam-multiple}
\end{figure}

Fig.~\ref{fig:rq2-exam-multiple} shows the cumulative distribution of EXAM scores comparing MBFL-DKMR against BLMu and Delta4Ms. The EXAM score distributions reveal that MBFL-DKMR consistently achieves lower examination effort compared to both state-of-the-art techniques. At an EXAM score of 0.02 (examining 2\% of the code), MBFL-DKMR successfully localizes over 15\% of the faults, while BLMu and Delta4Ms achieve slightly lower percentages. The curves consistently show that MBFL-DKMR requires less examination effort to locate the same percentage of faults compared to both advanced baseline techniques.
Furthermore, it can be observed that the performance advantage of MBFL-DKMR is consistency in the early part of the EXAM curve (EXAM < 0.10) which suggests that MBFL-DKMR is effective at ranking more faulty statements at the top of fault localization report compared to BLMu and Delta4Ms.

\begin{table}[htb]
  \centering
  \caption{Statistical Significance Analysis: MBFL-DKMR vs. state-of-the-art MBFL Techniques}
    \begin{tabular}{cllllc}
    \toprule
    \multicolumn{2}{c}{\multirow{2}[4]{*}{\textbf{Comparison}}} & \multicolumn{3}{c}{\textbf{p-value}} & \multirow{2}[4]{*}{\tabincell{c}{\textbf{Cliff's}\\
        \textbf{Delta}}} \\
\cmidrule{3-5}          &       & \multicolumn{1}{c}{\textbf{two-sided}} & \multicolumn{1}{c}{\tabincell{c}{\textbf{one-sided}\\
        \textbf{(less)}}} & \multicolumn{1}{c}{\tabincell{c}{\textbf{one-sided}\\
        \textbf{(greater)}}} &  \\
    \midrule

    \multirow{2}[0]{*}{MBFL-DKMR v.s.} & \quad BLMu & 2.51E-121 & 1.25E-121 & 1.00E+00 & -0.2067 \\
          & \quad Delta4Ms & 1.02E-84 & 5.11E-85 & 1.00E+00 & -0.0586 \\
    \bottomrule
    \end{tabular}
  \label{tab:rq2-significance-multiple}
\end{table}

We also conduct statistical significance testing using the Wilcoxon signed-rank test on the EXAM scores to validate the performance differences against state-of-the-art techniques. The results are presented in Table~\ref{tab:rq2-significance-multiple}. The statistical analysis confirms that MBFL-DKMR's performance improvements over both BLMu and Delta4Ms are statistically significant (p-value $<$ 0.05). The negative Cliff's Delta values indicate that MBFL-DKMR consistently achieves lower EXAM scores (better performance) compared to both advanced baselines. The effect size is medium for BLMu (-0.2067) and small for Delta4Ms (-0.0586), demonstrating that the improvements of MBFL-DKMR are not only significant but also practically meaningful.

In response to RQ2, our experimental results demonstrate that MBFL-DKMR achieves better performance compared to state-of-the-art MBFL techniques, establishing it as a competitive and effective approach for MBFL. MBFL-DKMR achieves improvements in Top-N localization accuracy, with 44 more faults localized at Top-1 compared to BLMu and 26 more compared to Delta4Ms. The MAP improvements of 4.1\% over BLMu and 4.3\% over Delta4Ms further confirm the ranking precision advantages of our approach. The EXAM score analysis reveals that MBFL-DKMR is particularly effective in the critical low-examination range, where developers typically focus their debugging efforts. The statistical significance of all improvements, with small to medium effect sizes, provides evidence that MBFL-DKMR's signal processing approach offers meaningful practical advantages over advanced techniques.

\subsection{RQ3: What is the time cost of MBFL-DKMR compared to existing MBFL techniques?}
\label{subsec:Results4RQ3}

RQ3 investigates the time cost of MBFL-DKMR compared to existing MBFL techniques. While the primary focus of MBFL-DKMR is to improve fault localization accuracy, it is also important to understand the computational overhead introduced by the DKMR component. For this research question, we measured the execution time of MBFL-DKMR and existing MBFL techniques across three key stages: Mutation Analysis (time required to generate mutants, execute tests, and construct the kill matrix), DKMR Processing (time required to apply signal enhancement and signal denoising to the kill matrix, only applicable to MBFL-DKMR), and Suspiciousness Calculation (time required to compute suspiciousness scores and rank statements).

\begin{table}[htb]
  \centering
  \caption{Time Cost Analysis of MBFL-DKMR}
    \begin{tabular}{lrr}
    \toprule
          \multicolumn{1}{c}{\textbf{MBFL-DKMR Stage}} & \multicolumn{1}{c}{\textbf{Avg. Time Cost (s)}} & \multicolumn{1}{c}{\textbf{Avg. Time Cost (\%)}} \\
    \midrule
    Mutation Analysis & 9644.18 & 99.9690\% \\
    DKMR Processing & 0.11  & 0.0011\% \\
    Suspiciousness Calculation & 2.88  & 0.0299\% \\
    \midrule
    Overall Time Cost & 9647.17 & 100.0000\% \\
    \bottomrule
    \end{tabular}%
  \label{tab:rq3-time}%
\end{table}%

Table~\ref{tab:rq3-time} presents the average execution time (in seconds) for each stage of the fault localization process. The results show that the mutation analysis stage dominates the overall execution time for MBFL-DKMR and existing MBFL techniques. This stage involves generating mutants, executing tests, and recording kill information, which are computationally intensive operations common to all MBFL techniques. The DKMR processing introduced by MBFL-DKMR adds an average of 0.11 seconds to the execution time, which represents less than 0.001\% of the total time. This overhead is negligible compared to the matrix construction stage and is a small price to pay for the improvements in fault localization accuracy demonstrated in RQ1-RQ2. The suspiciousness calculation stage is also efficient for both MBFL-DKMR and existing MBFL techniques, taking approximately 2.88 seconds on average. While it is slightly more time-consuming (2.88 seconds vs. 0.11 seconds) compared to the DKMR processing stage, but the difference is minimal.


In response to RQ3, our experimental results demonstrate that MBFL-DKMR introduces minimal computational overhead compared to existing MBFL techniques. The mutation analysis stage dominates the overall execution time for MBFL-DKMR, accounting for over 99.969\% of the total time. The DKMR processing introduced by MBFL-DKMR adds an average of 0.11 seconds to the execution time, which represents about 0.001\% of the total time. 
The negligible overhead of MBFL-DKMR's DKMR processing is a small price to pay for the improvements in fault localization accuracy. These results suggest that MBFL-DKMR's approach is not only effective at improving fault localization accuracy but also highly efficient and scalable.



\section{Threats to Validity}
\label{sec:ThreatstoValidity}

\noindent\textbf{Internal Validity.}
The first internal threat concerns the implementation of mutation tools, particularly the generation and execution of mutants. To mitigate this risk, we leverage established tools such as Major~\cite{just2014major} for traditional mutations. 
The second threat arises from the accuracy of our experimental setup for mutation analysis. We address this by aligning our test configurations with the Defects4J framework and validating results through manual verification~\cite{just2014defects4j}. 
Finally, the fidelity of the Fourier-based filtering process poses a potential threat due to its reliance on precise mathematical transforms and threshold settings. We mitigate this by using numerically stable libraries (e.g., NumPy) and validating the cutoff frequency $D_0$ through preliminary experiments.

\noindent\textbf{External Validity.}
The first external threat concerns the diversity of project scales and functionalities in our dataset. We use Defects4J v2.0.0~\cite{gay2020defects4j}, which includes 17 Java projects with real-world industrial faults and diverse characteristics, providing adequate representation across different software systems.
A second external threat is the dependency on specific suspiciousness formulas, which may limit the applicability of our results. We mitigate this by using a representative set of suspiciousness formulas (i.e., Dstar~\cite{wong2013dstar}, GP13~\cite{yoo2012evolving}, Jaccard~\cite{chen2002pinpoint}, Ochiai~\cite{abreu2006evaluation}, Op2~\cite{naish2011model}, and Tarantula~\cite{jones2005empirical}), ensuring robust comparisons across different MBFL approaches.

\noindent\textbf{Construct Validity.}
The primary construct threat relates to the evaluation metrics used in our study. While Top-N~\cite{zou2019empirical} and MAP~\cite{moffat2008rank} provide critical insights into fault localization effectiveness, they may not fully capture the practical aspects of debugging. To address this, we complement these metrics with EXAM~\cite{zou2019empirical,liu2018optimal}, which quantifies the effort required to locate faults. 
We also employ Wilcoxon signed-rank test~\cite{wilcoxon1945individual, liu2024delta4ms} and Cliff's Delta~\cite{macbeth2011cliff, amario2022understanding} effect size analysis to ensure statistical significance of our results.
Another potential construct threat is the inherent ordering of test cases and mutants in the kill matrix, which may introduce artificial periodicity or disrupt semantic correlations. 
We mitigate this by preserving code structure through mutant ordering based on source code positions and grouping test cases by test suites with alphabetical ordering within each suite rather than arbitrary criteria.

\section{Conclusion and Future Work}\label{sec:ConclusionandFutureWork}

In this paper, we propose DKMR (Denoising based Kill Matrix Refinement), a novel approach to address the noise issues in MBFL kill matrix through signal processing techniques. Our key insight is that the kill matrix can be viewed as signals containing meaningful fault-related patterns and high-frequency noise, where the noise significantly impacts localization accuracy. DKMR consists of two key stages: (1) signal enhancement through hybrid matrix construction that combines weak and strong kill information to improve SNR, making the denoising stage more effective, and (2) signal denoising through frequency domain filtering that suppresses noise while preserving fault-relevant patterns. Building on DKMR, we develop MBFL-DKMR (Mutation-Based Fault Localization with DKMR), a comprehensive fault localization framework that utilizes these refined matrices with fuzzy values for suspiciousness calculation, departing from traditional boolean-based approaches and providing more nuanced fault localization information.
Our evaluation on Defects4J v2.0.0 demonstrated that DKMR effectively suppresses noise and enhances fault-related signals in the kill matrix. Through ablation studies, we found that the denoising stage significantly improves fault localization effectiveness, while the signal enhancement stage further boosts performance by increasing the SNR to facilitate more effective denoising. When integrated into the MBFL framework, MBFL-DKMR achieves 129 faults localized at Top-1 position, significantly outperforming state-of-the-art techniques like BLMu (85 faults) and Delta4Ms (103 faults). The additional computational overhead introduced by our DKMR processing is only 0.11 seconds on average (0.001\% of total time), making MBFL-DKMR practical for real-world applications.

Future research could explore three promising directions:
(1) exploring better ordering strategies for mutants and test cases to improve the signal quality of the kill matrix;
(2) developing adaptive methods for automatically determining optimal parameters for the DKMR based on program characteristics;
(3) investigating the effectiveness of MBFL-DKMR under different SNR and extending the DKMR with more SNR robust signal processing techniques.


\newpage

\section{Data Availability}
The source code and experimental scripts are available through the conference submission system as supplemental material. The complete dataset cannot be shared due to large dataset size, but a representative sample is included in the replication package. Upon acceptance, the authors intend to make the data publicly available as requested.

\bibliographystyle{ACM-Reference-Format}
\bibliography{0-reference}

\end{document}